\documentclass{article}
\usepackage{spconf,amsmath,graphicx}
\usepackage{url,verbatim,subfigure,dsfont}
\usepackage{epsfig}
\usepackage{multirow}

\title{Audio Concept Classification with Hierarchical Deep Neural Networks}

\name{Mirco Ravanelli {\small $^{\star}$}, Benjamin Elizalde{\small $^{\star \dagger}$}, Karl Ni{\small $^{\dagger}$}\thanks{$^{\dagger}$Lawrence Livermore National Laboratory is operated by Lawrence Livermore National Security, LLC, for the U.S. Department of Energy, National Nuclear Security Administration under Contract DE-AC52-07NA27344.}, Gerald Friedland{\small $^{\star \dagger}$}\thanks{$^{\star \dagger}$Supported by the Intelligence Advanced Research Projects Activity (IARPA) via Department of Interior National Business Center contract number D11PC20066. The U.S. Government is authorized to reproduce and distribute reprints for Governmental purposes notwithstanding any copyright annotation thereon. The views and conclusion contained herein are those of the authors and should not be interpreted as necessarily representing the official policies or endorsement, either expressed or implied, of IARPA, DOI/NBC, or the U.S. Government.}}

\address{{\small $^{\star}$}Fondazione Bruno Kessler, Trento, Italy\\{\small $^{\star \dagger}$}International Computer Science Institute, Berkeley, USA\\{\small $^{\dagger}$}Lawrence Livermore National Laboratory, Livermore, USA}

\begin{document}
\ninept
\maketitle

\begin{abstract}
Audio-based multimedia retrieval tasks may identify semantic information in audio streams, i.e., audio concepts (such as music, laughter, or a revving engine). Conventional Gaussian-Mixture-Models have had some success in classifying a reduced set of audio concepts. However, multi-class classification can benefit from context window analysis and the discriminating power of deeper architectures. Although deep learning has shown promise in various applications such as speech and object recognition, it has not yet met the expectations for other fields such as audio concept classification. This paper explores, for the first time, the potential of deep learning in classifying audio concepts on User-Generated Content videos.  The proposed system is comprised of two cascaded neural networks in a hierarchical configuration to analyze the short- and long-term context information. Our system outperforms a GMM approach by a relative 54\%, a Neural Network by 33\%, and a Deep Neural Network by 12\% on the TRECVID-MED database.
\end{abstract}

\section{Introduction}

With the ubiquity of recording devices and online sharing websites, access to and the quantity of user-produced multimedia has grown exponentially. For this reason, recent competitive evaluations from NIST, i.e. the TRECVID Multimedia Event Detection (MED)~\cite{2012trecvidover}, and others like MediaEval and Pascal VOC, have focused on investigating core detection technologies to analyze, retrieve and label multimedia recordings based on their content.
The literature on audio concepts has been previously visited for various purposes including: laughter detection~\cite{ref1} and speech-music detection~\cite{ref2} just to mention a few. Nevertheless, most prior work has been based on test and training corpora recorded in laboratory, under rather controlled environment conditions, while the field of audio concept classification on User-Generated Videos (UGV) is relatively new. Identifying semantic information such as audio concepts (laughter, clapping, singing) is a significantly more complex problem on UGC datasets, due to multiple and sometimes overlap acoustic sources, variate durations and different level of prominence as well as different recording devices for each video. The authors in ~\cite{stephanieacm} train Gaussian Mixture Models (GMMs) for 20 acoustic concepts on User Generated Content (UGC) videos, but like ~\cite{zhengatech} evaluate their performance based on a higher level task, namely audio-based video event detection. In~\cite{kumar}, atomic units of sounds are employed to independently detect 10 concepts with over 50\% recall for all of them. Another example is~\cite{noisemes}, where GMMs were trained on the audio concepts segments. Then, the concepts were ordered in eight groups based on their average lengths. Results show an overall average segment classification accuracy of about 70\%. While GMMs techniques remain dominant, Neural Networks (NNs) based audio concept classification systems are still an underexplored research direction. 

Interest in NNs and deep architectures has recently been renewed due to several advancements in the vision and speech community. In particular,  the introduction of pre-training methods such as the unsupervised, greedy layer-wise technique based on Restricted Boltzmann Machines (RBM)~\cite{ref29}, has made the training of networks with many hidden layers feasible and effective. The RBM technique has enabled demonstrable improvements in character recognition \cite{ref15}, 
object recognition \cite{ref17}, information retrieval \cite{ref19}, and 
speech recognition \cite{ref21},
just to name a few. 
As a consequence, Deep Neural Networks (DNN) and more recently Hierarchical Deep Neural Networks (H-DNN) have been proven to be effective for several applications, including speech recognition \cite{ref22,ref24}. The success  of H-DNN is attributed to the intrinsic capabilities of deep learning, but more importantly, to the analysis of the short and long term audio modulation. Such acoustic information is based on ingesting the short and long-term context windows to the NNs. 

In this paper, we explore H-DNN for acoustic concept classification and present its contribution to the field. Our approach is partially inspired by the H-DNN paradigm recently adopted for domain specific speech recognition \cite{ref26}. Although H-DNN had promising results in the speech field, there was no indication of a similar performance in our UGC task. We investigate the capabilities and limitations of H-DNN's in the context of the UGC application. Specifically, our H-DNN approach relies on short- and long-term cascade analysis of audio concepts, not included in a standard DNN. Because H-DNN's output are probability posteriors, they can be used as semantic-low-level features for multimedia retrieval systems. In particular, in \cite{slam2014} we show that these features can indeed be used to show evidence of concept occurrence on high-level semantic scenes such as a wedding ceremony, a soccer game or a broadcast news. 

The remainder of this paper addresses these contributions and is broken up as follows. The proposed deep learning architecture is discussed in Sec.~\ref{sec:dl-acc}. Sec.~\ref{sec:setup} describes the experimental setup, including a description of the corpora. Final results are presented in Sec.~\ref{subsec:accres}, and Sec.~\ref{sec:conclusion} concludes the work.



\section{Deep Learning of Audio Concepts} \label{sec:dl-acc}



\begin{figure}[t!]
   \centering
     \includegraphics[width=0.5\textwidth]{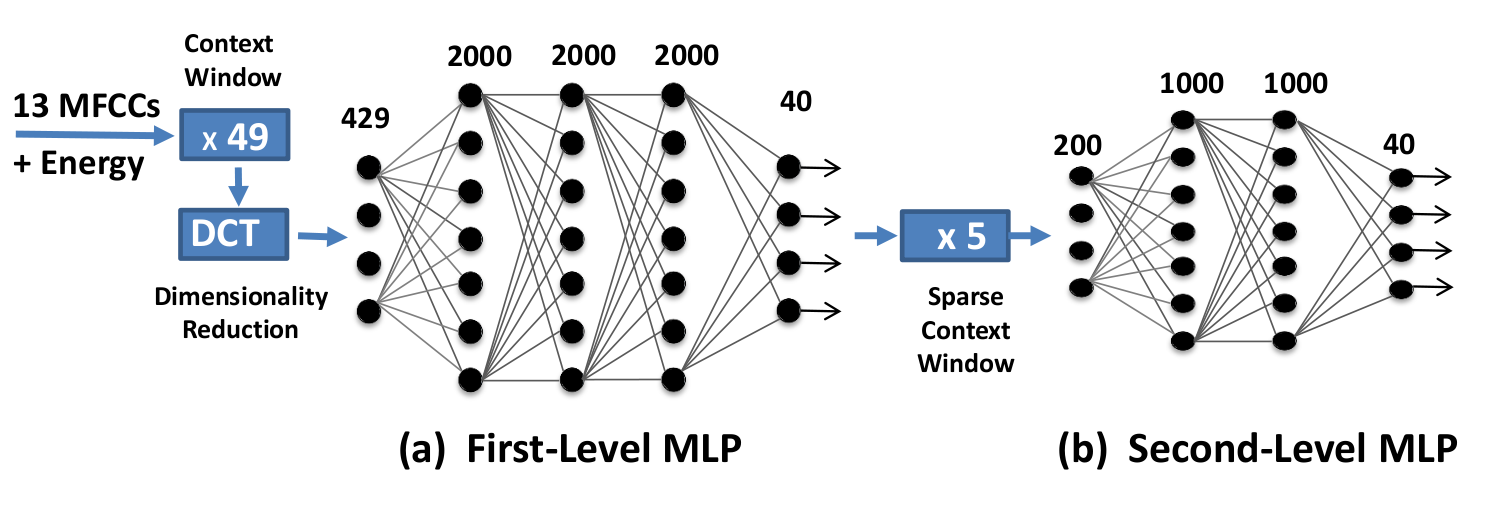}
     \caption{The proposed Hierarchical Deep Neural Network (H-DNN) architecture for audio concept classification.}
     \label{fig:img1}
\end{figure}

Short- and long-term modulations of the H-DNN model are extremely effective for speech recognition and the same modeling could benefit audio concept classification. Most audio concepts such as music, clapping, knocking, laughing  and many others are often characterized by several replicas of similar pattern over the time, thus suggesting that a long-term analysis of the acoustic concepts could be convenient. Our specific implementation calls for two NNs that are incorporated as a cascade into a single H-DNN architecture. Lastly, using sigmoid-based neurons and a softmax classifier for the output layer, the neural network, unlike multi-class SVM-based systems, generates an output set of posterior probabilities. The proposed two-fold MLP system is depicted in Fig.~\ref{fig:img1}.

\subsection{Input Features}
\label{sec:input}
The extracted acoustic features are the Mel-Frequency Cepstral Coefficients (MFCCs) C0-C12, with energy included, for a total of 14 dimensions, typical for systems like our proposed architecture. Each feature frame is computed using a 25 ms Hamming window with a stride size of 10ms per frame shift. After a mean and variance normalization step, we apply a context window that gathers several consecutive frames. The importance of the role of the context window is reported in Sec.~\ref{subsec:accres}, showing significant benefits in its usage. The context window processes 49 consecutive frames (centered at the $25^{th}$ frame). Finally, prior to the input layer of the first NN, a dimensionality reduction step using the Discrete Cosine Transform (DCT) is applied to de-correlate the information in time. The final NN input dimensionality consists of 429 features. 

\subsection{Hierarchical Processing} \label{sec:hp}
The first level NN in Fig.~\ref{fig:img1}(a), fed by the input stream described in Sec.~\ref{sec:input}, converts the low-level features into a higher level representation. This NN is composed of 3 hidden layers with 2000 neurons per layer. Because this NN consists of more than 2 hidden layers, pre-training based on the RBM is important for initialization and convergence purposes. The output layer, a softmax-based classifier, outputs a 40 dimensional vector, corresponding to the number of audio concepts provided for classification. Examples of audio concepts are laughing, clapping, cheering, speech, music and singing. The full list of the adopted audio concept is reported in ~\cite{ref13}.

Afterwards, additional processing on the features generated by the first NN (Fig.~\ref{fig:img1}(a)) is conducted prior to feeding into the second NN. The long term analysis of the audio concepts is achieved by further selecting several sparse input frames. In this case, we sample the features generated by the NN in  Fig.~\ref{fig:img1}(a) at the 5 positions: -10, -5, 0, +5,+10. Hence, the total in input dimensionality of Fig.~\ref{fig:img1}(b) is 200.

The long-term NN Fig.~\ref{fig:img1}(b) is composed of 2 hidden layers with 1000 neurons each. Compared to Fig.~\ref{fig:img1}(a), the NN in Fig.~\ref{fig:img1}(b) employs a shallower architecture. We justify the implementation choice due to the simpler task it has to perform. That is, the MLP in Fig.~\ref{fig:img1}(a) realizes a conversion from low-level to high level features, while Fig.~\ref{fig:img1}(b) operates on already processed input streams from the previously trained NN.

\subsection{Restricted Boltzmann Machine Pre-Training}
Pre-training replaces random initialization of the parameters with a justified and more convenient weight initialization, without which it is usually difficult to employ more than one or two hidden layers using back-propagation training. A number of pre-training techniques have been explored previously, including both discriminative approaches \cite{IEEEexample:rbm4} and unsupervised methods based on a stack of Restricted Boltzmann Machines \cite{IEEEexample:rbm1}. We have chosen to use the unsupervised approach, which was successfully used to improve speech recognition performance for TANDEM \cite{IEEEexample:rbm3}, bottleneck \cite{IEEEexample:rbm2}, and hierarchical bottleneck approaches~\cite{ref24} 
as well as for hybrid speech recognition \cite{IEEEexample:intro1}.

Next, we trained each adjacent pair of NN layers as an RBM in an efficient greedy, layer-wise technique, initializing the NN weights. The derived weights are directly used to initialize the MLP and, by means of fine-tuning phase carried out using the standard back-propagation algorithm, a joint optimization of all the layers is performed. 
We take advantage of RBM pre-training only over the first NN in Fig.~\ref{fig:img1}(a), while a random initialization is performed for the second NN in Fig.~\ref{fig:img1}(b). This choice is connected to the decision of using a deeper MLP for first level NN as discussed in \ref{sec:hp}. Since pre-training appears to be useful for NNs with many hidden layers, no significant improvement could potentially be obtained exploiting pre-training on the shallower MLP.

\section{Experimental Setup}
\label{sec:setup}

This section describes the video corpora used for the audio concepts classification in Sec.~\ref{subsec:corpora}, the GMM and NN training algorithms in Sec.~\ref{subsec:mlp} and the evaluation metric in Sec.~\ref{subsec:evaluation}.

\subsection{Corpora Description}
\label{subsec:corpora}

The audio concepts belong to the TRECVID MED 2012 dataset \cite{2012trecvidover}, which contains UGC videos of about three minutes each. The audio from the videos contains environmental acoustics, overlapped sounds and unintelligible audio among other characteristics. The manually created annotations are based on three different sources. First, SRI-Sarnoff~\cite{sarnoffconcepts} set consists of 28 concepts from 291 videos for a total of 11.6 hours. Second, CMU~\cite{noisemes} set consists of 42 concepts from 216 videos taken from MED 2012, totaling 5.6 hours. Lastly, Stanford~\cite{stephanieacm} set consists of 20 concepts from 1138 videos, totaling 11.87 hours. The audio concepts are audio trimmed from the main recording based on the annotation. Concepts have variable length. 

For the experiments described in the paper, we took the total 90 concepts and used 80\% of the annotations to train and 20\% to test the NN described in Sec.~\ref{subsec:baseline} as the baseline, using per-frame concept accuracy as the evaluation metric. Afterwards, we chose the top 40 concepts with highest accuracy. The reason for the cut off was due to accuracies been close to zero for the rest of the concepts. This procedure is described in more detail in~\cite{ref13}.

\subsection{Neural Network \& GMM Systems Training}
\label{subsec:mlp}

The NN training and pre-training phases are based on the GPU version of the TNet toolkit \cite{ref30}. Pre-training initializes weights in the first two hidden layers via RBM (Gaussian-Bernoulli) using a learning rate of 0.005 with 10 pre-training epochs. The remaining RBMs (Bernoulli- Bernoulli) use a learning rate of 0.05 with 5 pre-training epochs. From the training-set, a small cross-validation set (10\% of training data) has been derived for the following back-propagation training. The fine-tuning phase is performed by a stochastic gradient descent optimizing cross-entropy loss function. The learning rate, is kept fixed at 0.002 as long as the single epoch increment in cross-validation frame accuracy is higher than 0.5\%. For subsequent epochs, the learning rate is halved until the cross-validation increment of the accuracy is less than the stopping threshold of 0.1\%. NN weights and biases are updated per blocks of 1024 frames.

A conventional GMM approach is also used to provide a baseline comparison with our H-DNN results. The system has two steps: the creation of the concepts models (training), and the scoring (testing). In the first step the Expectation Maximization (EM) algorithm updates a pre-trained concept-independent GMM, known as the Universal Background Model (UBM), to create a 256-mixture GMM for each audio concept. In the second step, a log-likelihood ratio is used to obtain a similarity score between each concept-dependent GMM and the acoustic features of each test audio. The UBM is used in the likelihood-ratio computation for score normalization. 

\subsection{Evaluation Metric}
\label{subsec:evaluation}

Results are quantitatively analyzed with classification frame accuracy (F.A.), which is evaluated by comparing the label from the frame's highest posterior against its corresponding ground truth label.

\section{Experimental Results} 
\label{subsec:accres}

This section provides the quantitative justification for using H-DNNs by using metrics defined in Sec.~\ref{sec:setup}. Empirical results include relative performance by sweeping parameters and comparisons to baseline systems. Sec.~\ref{subsec:baseline} reports the baselines performance, Sec.~\ref{subsec:context} explores the role of contextual information while Sec.~\ref{subsec:optarch} explores multiple architectural considerations in the NN implementation. Finally, Sec.~\ref{subsec:hierarchical} evaluates the entire H-DNN system (including long-term NN, Fig.~\ref{fig:img1}(b)) by comparing to baselines.

\subsection{Baseline Performance} \label{subsec:baseline}
To compare the proposed approach, we have employed two conventional GMM baselines. The first one is based on MFCCs+$\Delta$+$\Delta\Delta$ coefficients, while the second system is based on the MFCCs described in Sec.~\ref{sec:input}, but adopting context windows of five consecutive frames. Both the number of iterations and the number of gaussians are optimized over the cross-validation set. To add to the comparison, a shallow NN baseline is also presented. The NN architecture is composed by a single  hidden layer of 1000 neurons, and then fed by the MFCC coefficients with a context window of nine consecutive frames. The baseline performances are summarized in Table \ref{tab:baselines}.

\begin{table}[ht]
\centering
\begin{tabular}{| c | c | c | c | c | c | c |}
  \hline
    \bfseries System  & \bfseries Features    & \bfseries F.A.(\%)  \\
  \hline
   GMM Baseline 1 & 42: 14 MFCCs+$\Delta$+$\Delta\Delta$   & 23.52 \\ 
  \hline
   GMM Baseline 2  & 70: 14 * 5 MFCCs  & 24.07   \\
  \hline
   NN Baseline   & 126:14 * 9 MFCCs & 27.70  \\
  \hline
\end{tabular}
\caption{Audio concepts per-frame classification performance (F.A.\%) for the baseline systems. The GMMs systems used 256 gaussians, while the NN baseline is composed of a single hidden layer of 1000 neurons.}
 \label{tab:baselines}
\end{table}

The shallow NN system has a F.A.(\%) of 27.70\%, while the best GMM baseline performance is 24.07\%. As discussed in Sec.~\ref{subsec:context}, the performance of the NN is significantly better than the GMM, mainly due to a better use of the context information (Fig. \ref{fig:img2}).







\subsection{Role of the context windows} 
\label{subsec:context}

\begin{figure}[t]
   \centering
     \includegraphics[width=0.45\textwidth]{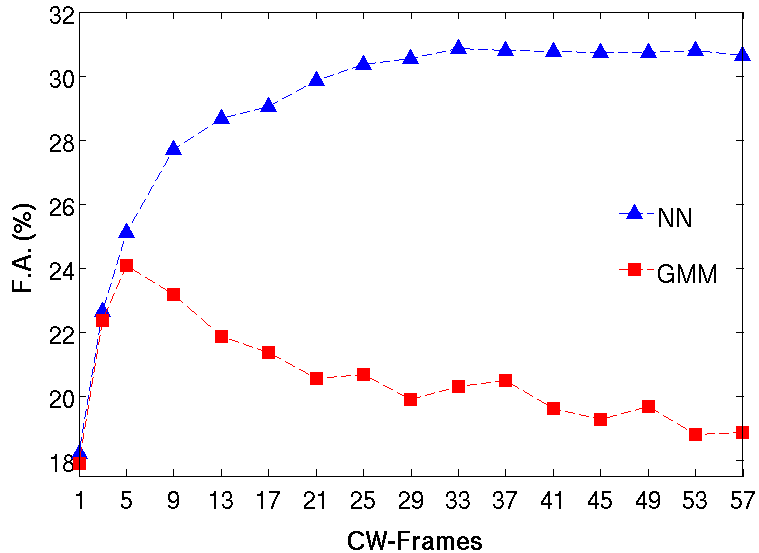}
     \caption{Per-frame classification accuracy (\%) progression by increasing the context window. The blue markers refer to the NN system, while the red squares refer to the GMM baseline.}
     \label{fig:img2}
\end{figure}

In Fig.~\ref{fig:img2}, starting from the same NN architecture of the baseline (one hidden layer of 1000 neurons), we show the result of sweeping various context window sizes. As expected, for the NN case, there is a consistent improvement for progressively larger context windows. The best performance has been obtained when a context windows of 33 frames (which correspond to a context of 345 ms with an NN input dimensionality of 462) is adopted. This setting lead to a per-frame accuracy percentage of 30.87\% corresponding to a relative improvement of over 28\% above the no-context window case. Meanwhile, the NN baseline is outperformed by up to a relative 11\%. The Fig. \ref{fig:img2} also shows a logarithmic behaviour, consisting of sharp initial improvements in performance by adding the closest frames followed by a saturation. Typical reasons for slowed or discontinued performance improvement can be ascribed to both, the deficit of useful information at temporally distant times and the large dimensionality problems. For the sake of comparison, the sweeping of the context windows has been applied for also the GMM baseline. The Fig. \ref{fig:img2} clearly shows that increasing the context windows of more than five consecutive frames lead to a decrease of performance. This trend suggests that the GMM paradigm is not fully adequate in classifying high-dimensionality vectors.

To improve the context description for the NN case, we further extend the capability of the context window by adopting a DCT-based dimensionality reduction after a longer frame windowing. Such a transformation also serves to decorrelate the information inside the context windows. Table \ref{tab:tab3} compares various context window sizes while fixing the input dimensionality of the NN at 462.

\begin{table}[ht]
\centering
\begin{tabular}{| c | c | c | c | c | c | c |}
  \cline{1-7}
    \bfseries   & \bfseries x33  & \bfseries x37   & \bfseries x41 & \bfseries x45 & \bfseries x49 & \bfseries x53 \\
  \hline
   NO DCT & 30.87 & 30.81  & 30.78 & 30.75 & 30.80 & 30.72  \\
  \hline
   DCT  & 31.01 & 31.39  & 32.00 & 32.24 & \bfseries 32.30 & 32.27  \\
  \hline

\end{tabular}
\caption{The role of the DCT based dimensionality reduction. For the DCT case, the input dimensionality of the NN is kept fixed at 462, while the context window which is applied to the input features is swept from 33 to 53 consecutive frames.}
 \label{tab:tab3}
\end{table}

The improvement of larger context windows in audio concept classification is explicit and the general trend is encouraging. In particular, exploiting a context window of 49 frames (which corresponds to about half second of time context), we have improved the NN baseline of more than 16\%. The result is consistent with similar results obtained for TRAPs~\cite{traps} and HATs~\cite{hats} systems, where a 500 ms context information has been exploited to improve speech recognition performance. No improvements has been verified by applying the DCT-based dimensionality reduction to the GMM baseline.

\subsection{Architecture Optimization} \label{subsec:optarch}

Table~\ref{tab:tab4} explores modifications to the system described in Sec.~\ref{subsec:context}. The modifications of the neural network architecture include both number of hidden layers and number of neurons per hidden layer. We also studied the impact of RBM pre-training on the audio concept classification task.

Results show a consistent improvement when more than one hidden layer has been adopted. Note the significant performance improvement going from one to two hidden layers. Interestingly, for our application of concept classification, no substantial improvement arises from employing more than 2 hidden layers. Also a remarkable improvement occurs when more than 500 neurons are used per hidden layer. Pre-training, as expected, seems useful only in the presence of deep and wide architectures, where a large quantity of parameters have to be estimated. The best performance is obtained when a network with three hidden layers and 2000 neurons each is pre-trained with RBM.

\begin{table}[ht]
\centering
\begin{tabular}{| c | c | c | c | c | c | c | }
  \hline
  \multirow{2}*{} & \multicolumn{2}{| c |}{500 neurons} & \multicolumn{2}{| c |}{1000 neurons}  & \multicolumn{2}{| c |}{2000 neurons} \\
  \cline{2-7}
  & RND & RBM & RND & RBM & RND & RBM \\
  \hline
  1 Layer & 29.16 & 29.14 & 30.61 & 30.65  & 30.81 & 30.85\\
  \hline
  2 Layers& 31.00 & 31.11 & 32.30 & 32.33  & 32.80 & 32.92\\
  \hline
  3 Layers& 31.14 & 31.15 & 32.62 & 32.73  & 32.68 & \bfseries 32.96\\
  \hline
  4 Layers& 31.05 & 31.25 & 32.29 & 32.49  & 30.20 & 32.57\\
  \hline
\end{tabular}
\caption{Optimization of the architecture of NN in Fig.~\ref{fig:img1}(a), the primary neural network. Rows report the number of hidden layers, while columns refer to the number of neurons for each hidden layer. RND shows the per-frame accuracy performance achieved with a random initialization of the weights, while RBM takes advantage of pre-training.}
 \label{tab:tab4}
\end{table}

\subsection{Hierarchical Processing} 
\label{subsec:hierarchical}
Table~\ref{tab:hierarchical} depicts classification performance under various techniques, defined for each row. The baseline systems introduced in Sec.~\ref{subsec:baseline} corresponds to the first two rows. The third row, contains the NN system of Sec.\ref{subsec:context}, with a DCT-compacted and 49 consecutive frame context window. Next is the deeper (three hidden layer) and wider (2000 neurons per layer) DNN architecture with RBM pre-training described in Sec.\ref{subsec:optarch} and depicted in Fig. \ref{fig:img1}(a) . Lastly, the H-DNN uses the DNN as its foundation and is incorporated into the hierarchical processing introduced in Sec.\ref{sec:hp} and depicted in Fig. \ref{fig:img1}(a)(b). 

Besides the improved performance by using a large context window, our initial hypothesis of benefiting from a long-term analysis of the audio concepts is realized. Indeed, audio concepts, such as music, clapping, knocking, etc. are often characterized by several replicas of similar pattern over the time, validating the ability of long-term analysis to better describe such acoustic events. In fact, the improvement of such a system is a relative 12\% over the best DNN described in Sec.~\ref{subsec:optarch}.

\begin{table}[ht]
\centering
\begin{tabular}{| c | c | c | c | c | c | c |}
  \hline
    \bfseries System  & \bfseries CW  & \bfseries DCT   & \bfseries RBM & \bfseries HP & \bfseries F.A.(\%)  \\
  \hline
   Baseline GMM & x5 & no  & no & no & 24.07  \\ 
  \hline
   Baseline NN  & x9 & no  & no & no & 27.70  \\
  \hline
   DNN  & x49 & yes & yes & no & 32.96  \\
  \hline
   H-DNN  & x49 & yes & yes & yes & \bfseries 36.93  \\
  \hline
\end{tabular}
\caption{Audio concepts per-frame classification performance (F.A.\%). The first row refers to the GMM baseline, while the rest of the rows are obtained by progressively improving the NN baseline. The  CW column reports the adopted context window, while DCT refers to the presence of the DCT-based dimensionality reduction. The column RBM refers to the pre-training with RBM , and  finally,  HP reports whether the Hierarchical Processing performed by the second MLP is enabled.}
 \label{tab:hierarchical}
\end{table}
Some examples of the concept classification performance comparison between the H-DNN and the baseline NN are: wind (Stanford) 63\%-51\%, speech (CMU) 68\%-58\% and metallic noises (SRI) 73\%-53\%, mumble (CMU) 13\%-10\%, bird (CMU) 13\%-5\% and quite engine (CMU) 6\%-3\%. 
Since the H-DNN system is composed of 7 hidden layers, for the sake of comparison a DNN with the same number of hidden layers and with the same number of neurons should be proposed. Unfortunately, such a comparison is not feasible since a single DNN with this architecture gets stuck in a poor local minima during the training phase. This can be due to the large number of parameters to determine, compared to the  size of the training corpus (curse of dimensionality problem). The H-DNN, which is based on two different MLPs trained independently, is more robust against this problem, allowing us to employ several processing layers for classifying the acoustic concepts.
 

\section{Conclusions}
\label{sec:conclusion}
This paper explores for the first time the advantage of deeper architectures for classifying audio concepts in audio from UGC videos. The proposed system employs a H-DNN of two cascaded neural networks, which successfully explores both short-term modulation, through a context window of the first neural network, and long-term modulation, through the sparse context window of the H-DNN. Moreover, H-DNN significantly outperforms Gaussian Mixture Model and Neural Network baselines, as well as Deep Neural Network-based classification systems. Our research suggests promising results for deep architectures on audio concepts and immediate future work is currently being conducted on the analysis of audio concept posteriors as semantic features for an audio-based video event detection system.

{\small
\bibliographystyle{IEEEbib}
\bibliography{main}
}

\end{document}